\documentclass[twoside,twocolumn]{article}
\usepackage{times}
\usepackage{latexsym}
\usepackage{amsbsy}
\usepackage{ifthen}
\usepackage{archiv2e}

\usepackage[dvips]{epsfig}

\begin{document}
%
% ************ To be filled in by the Editor *************
\Eingangsdatum{December xx, 2005}  % . Revised Month 00, 2001}
\Kolumnentitel{
A. Lakhtakia, J.B. Geddes III:
Scattering by nihility cylinder}
%\Kolumnentitel{Letters}
\Autoren{\Name{f.}{vv}{ll}{j.}} \Sachgebiet{1} \Band{60}
\Jahr{2006} \Heft{11} \Ersteseite{1}
%
% Der folgende Befehl wird nur bei 'Paper' benoetigt,
% er entfaellt bei 'Letter' (auskommentieren).
\Letzteseite{3}
%\Widmung{}
%\Leerspalte
%\Vorschub{}
%
% ********************** Start here! ****************************
%
% ****************** Author defined macros *********************
% If applicable, define your own macros  by the command \newcommand
% right after here.

%%% personal macros - start
%%% please no definitions, font changes, size changes
%%% within the text

% for 3-vectors/dyadics
\def\##1{{\bf{#1}}}
\def\=#1{\underline{\underline{#1}}}

% for 6-vectors/dyadics

\def\+#1{\underline{\bf #1}}
\def\*#1{\underline{\underline{\bf #1}}}

\def\eps{\epsilon}
\def\epso{\epsilon_0}
\def\muo{\mu_0}
\def\ko{k_0}
\def\co{c_{\scriptscriptstyle 0}}
\def\lambdao{\lambda_0}
\def\etao{\eta_0}
\def\.{\mbox{ \tiny{$^\bullet$} }}

\def\curl{\nabla\times}
\def\div{\nabla \mbox{ \tiny{$^\bullet$} }}

\def\id{\#{\#{I}}}

\def\ux{\#{\hat u}_x}
\def\uy{\#{\hat u}_y}
\def\uz{\#{\hat u}_z}

\def\le{\left(}
\def\ri{\right)}
\def\les{\left[}
\def\ris{\right]}
\def\lec{\left\{}
\def\ric{\right\}}

\def\c#1{\cite{#1}}
\def\l#1{\label{#1}}
\def\r#1{(\ref{#1})}

\def\epsr{\epsilon_r}
\def\mur{\mu_r}

\newcommand{\mb}[1]{\mbox{\boldmath$\bf#1$}}

%%% personal macros - end

% ****************  Begin of manuscript **********************
% ************* Opening part of the manuscript ***************
% Fill in between the braces of the commands:

% ----- Key words:

% ----- Title of the Paper:
\Title{Scattering by a Nihility Cylinder}

% ----- Authors' names:
% full first name, initials, lastname
\Authors{Akhlesh Lakhtakia, Joseph B. Geddes III}

% ----- Addresses:
% academic titles, initials and names of authors,
% department, institution, street, P.O. Box,
% city, zip code, country, e-mail.
\Addresses{A. Lakhtakia,
 CATMAS---Computational \& Theoretical
Materials Science Group, Department of Engineering Science and
Mechanics, Pennsylvania State University, University Park, PA
16802--6812, USA. E--mail: akhlesh@psu.edu\\
J.B. Geddes III,
 CATMAS---Computational \& Theoretical
Materials Science Group, Department of Engineering Science and
Mechanics, Pennsylvania State University, University Park, PA
16802--6812, USA. E--mail: jbg136@psu.edu
}

% ----- Abstract:
\Abstract{The total scattering and the extinction efficiencies of a nihility
cylinder of infinite length and circular cross--section are identical and independent
of the polarization state of a normally incident plane wave.}

% Ignore the following command,
% it will be activated by the Editor if appropriate
% (Erzeugt Rubriktitel 'Letters')
% \Rubrik{}{0mm}{0mm}

% ----- Paper or Letter:
% Choose appropriately
\Letter
%\Paper

% ************* Begin of the body of the manuscript **********

\section{Introduction}

The emergence of nihility as an electromagnetic {\em medium}   \c{ET} can be attributed to
the rather extraordinary developments on negatively refracting materials during
this decade \c{LMW_aeu, Rama}. Much of the impetus for this development
has been provided by the prospect of the so--called perfect lens \c{Pen}. Any perfect
lens in the present context is required to simulate nihility \c{PLN1,PLN2}.

The relative permittivity  and the relative permeability  of   nihility are 
null--valued.  Clearly,
nihility is unachievable, but it may be approximately simulated in some narrow
frequency range~---~hence, its attraction \c{Ziol,LM05}. Reflection
and refraction of plane waves due to nihility half--spaces \c{LM05} and slabs \c{PLN1,PLN2}
has been studied in some detail, as well as the scattering of plane waves by nihility
spheres \c{LL06}.
Along the same lines, this communication focuses on the canonical problem \c{vdB,BH}
of the  scattering
response of a nihility cylinder of circular cross--section and infinite length to a normally
incident plane wave.
An $\exp(-i\omega t)$ time--dependence is implicit
in the following sections.

\section{Boundary Value Problem}
The geometry of the canonical problem is best stated using the cylindrical
coordinate system $(\rho,\phi,z)$. 
The cylinder $\rho\leq a$ is oriented parallel to the $z$ axis; and, with its wave vector
parallel to the $-x$ axis, a plane wave is normally
incident on this cylinder. 
Two different cases must be considered: (i) The incident magnetic field phasor is parallel
to the $z$ axis, (ii) the incident electric field phasor is parallel to the $z$ axis. 

Direct derivation for a nihility cylinder being evidently intractable, a limiting procedure has
to be resorted to. Therefore, let
the relative permittivity of the cylinder be denoted by $\eps_r$,
and its relative permeability by $\mu_r$. The standard results for this
cylinder \c{vdB,BH} can be manipulated for nihility cylinders.

\subsection{Case (i)} 
The incident field exists everywhere in the region of interest when the
scatterer is absent.
Therefore, the incident electric and magnetic field phasors may be stated as
follows \c{BH}:
\begin{eqnarray}
&&
\left.\begin{array}{l}
\#E_{inc}(\rho,\phi) = \sum_{n=-\infty}^\infty\,\alpha_n\,
\#M_n^{(1)}(\rho,\phi;\ko)\\[5pt]
\#H_{inc}(\rho,\phi) = (i\etao)^{-1}\,\\
\qquad\quad\times\sum_{n=-\infty}^\infty\,\alpha_n\,
\#N_n^{(1)}(\rho,\phi;\ko)
\end{array}\right\}\,,
\nonumber
\\
&&\qquad\qquad\qquad \rho\geq 0\,,
\end{eqnarray}
where $\ko$ is the wavenumber in and $\etao$ is the intrinsic impedance of 
free space (i.e., vacuum), while the coefficients
\begin{equation}
\alpha_n=(-i)^{n+1}/\ko\,.
\end{equation}

The wavefunctions used in the foregoing equations and hereafter are defined 
as follows:
\begin{eqnarray}
\nonumber
\#M_n^{(1)}(\rho,\phi;k) &=& k\left(in \frac{J_n(k\rho)}{k\rho}\,\hat{\rho} \right.\\[4pt]
&&\qquad \left.- \frac{dJ_n(k\rho)}{dk\rho}\,\hat{\phi}
\right)\,e^{in\phi}\,,\\[5pt]
\#N_n^{(1)}(\rho,\phi;k) &=& k \,J_n(k\rho)\, 
e^{in\phi}\,\hat{z}\,,\\[5pt]
\nonumber
\#M_n^{(3)}(\rho,\phi;k) &=& k\left(in \frac{H_n^{(1)}(k\rho)}{k\rho}\,\hat{\rho} \right.\\[4pt]
&& \qquad\left.- 
\frac{dH_n^{(1)}(k\rho)}{dk\rho}\,\hat{\phi}
\right)\,e^{in\phi}\,,\\[5pt]
\#N_n^{(3)}(\rho,\phi;k) &=& k \,H_n^{(1)}(k\rho)
\,e^{in\phi}\,\hat{z}\,.
\end{eqnarray}
Whereas $J_n(\xi)$ are Bessel functions, $H_n^{(1)}(\xi)$ are Hankel functions
of the first kind, with $\xi$ denoting the argument.

The scattered field phasors are given by
\begin{eqnarray}
&&
\left.\begin{array}{l}
\#E_{sca}(\rho,\phi) = \sum_{n=-\infty}^\infty\,\alpha_n\,a_n\,
\#M_n^{(3)}(\rho,\phi;\ko)\\[5pt]
\#H_{sca}(\rho,\phi) = (i\etao)^{-1}\,\\
\qquad\quad\times\sum_{n=-\infty}^\infty\,\alpha_n\,a_n\,
\#N_n^{(3)}(\rho,\phi;\ko)
\end{array}\right\}\,,
\nonumber
\\
&&\qquad\qquad\qquad \rho\geq a\,,
\end{eqnarray}
where
\begin{eqnarray}
\nonumber
&&
a_n=-\les \eta_r\,J_n(\ko a) - J_n^\prime(\ko a) L_n(\ko n_r a)\ris\\
&&\qquad
\les \eta_r\,H_n^{(1)}(\ko a) - H_n^{(1)\prime}(\ko a) L_n(\ko n_r a)\ris^{-1}\,,
\label{an}
\end{eqnarray}
the prime denotes differentiation with respect to the argument,
the functions
\begin{equation}
L_n(\xi)= \frac{J_n(\xi)}{J_n^\prime(\xi)}\,,
\end{equation}
the relative impedance $\eta_r=\sqrt{\mu_r/\epsilon_r}$, and the refractive
index $n_r=\sqrt{\epsr\mur}$.

The total scattering efficiency is the sum
\begin{equation}
\label{Qsi}
Q_{sca}^{(i)} = \frac{2}{\ko a}\,\sum_{n=-\infty}^{\infty}\, \vert a_n\vert^2\,,
\end{equation}
and the extinction efficiency may be derived from the optical theorem \cite[Sec. 3.4]{BH}
as
\begin{equation}
\label{Qei}
Q_{ext}^{(i)}=-
\frac{2}{\ko a}\,\Re\,\Big(\sum_{n=-\infty}^{\infty}\, a_n\Big)\,,
\end{equation}
where $\Re$ stands for `the real part of'.

\subsection{Case (ii)} 
The incident electric and magnetic field phasors may be stated as
follows \c{BH}:
\begin{eqnarray}
&&\left.\begin{array}{l}
\#E_{inc}(\rho,\phi) = \sum_{n=-\infty}^\infty\,\beta_n\,
\#N_n^{(1)}(\rho,\phi;\ko)\\[5pt]
\#H_{inc}(\rho,\phi) = (i\etao)^{-1}\,\\
\qquad\quad\times\sum_{n=-\infty}^\infty\,\beta_n\,
\#M_n^{(1)}(\rho,\phi;\ko)
\end{array}\right\}\,,
\nonumber
\\
&&\qquad\qquad\qquad
 \rho\geq 0\,,
\end{eqnarray}
where 
\begin{equation}
\beta_n=i\alpha_n\,.
\end{equation}

The scattered field phasors are given by
\begin{eqnarray}
&&
\left.\begin{array}{l}
\#E_{sca}(\rho,\phi) = \sum_{n=-\infty}^\infty\,\beta_n\,b_n\,
\#N_n^{(3)}(\rho,\phi;\ko)\\[5pt]
\#H_{sca}(\rho,\phi) = (i\etao)^{-1}\,\\
\qquad\quad\times\sum_{n=-\infty}^\infty\,\beta_n\,b_n\,
\#M_n^{(3)}(\rho,\phi;\ko)
\end{array}\right\}\,,
\nonumber
\\
&&\qquad\qquad \qquad\rho\geq a\,,
\end{eqnarray}
where
\begin{eqnarray}
\nonumber
&&
b_n=-\les J_n(\ko a) -\eta_r\, J_n^\prime(\ko a) L_n(\ko n_r a)\ris\\
&&\qquad
\les  H_n^{(1)}(\ko a) - \eta_r\,H_n^{(1)\prime}(\ko a) L_n(\ko n_r a)\ris^{-1}\,.
\label{bn}
\end{eqnarray}

The total scattering efficiency and the extinction efficiency, respectively, are as follows:
\begin{eqnarray}
\label{Qsii}
&&Q_{sca}^{(ii)} = \frac{2}{\ko a}\,\sum_{n=-\infty}^{\infty}\, \vert b_n\vert^2\,,
\\
\label{Qeii}
&&Q_{ext}^{(ii)}=-
\frac{2}{\ko a}\,\Re\,\Big(\sum_{n=-\infty}^{\infty}\, b_n\Big)\,.
\end{eqnarray}

\subsection{Limiting Procedure for Nihility Cylinder}\label{spec}
Now, the refractive index of nihility must be null--valued because $\epsr=\mur=0$.
 For the functions $L_n(\xi)$,
we have 
\begin{eqnarray}
&&
\lim_{\xi\to 0} \xi L_0(\xi) = -2\,,\\[5pt]
&&
\lim_{\xi\to 0} \xi^{-1} L_n(\xi) = n^{-1}\,,\quad n\ne 0\,.
\end{eqnarray}
Therefore, after taking the limit $n_r\to 0$, \r{an} and \r{bn} for a nihility cylinder simplify to
\begin{eqnarray}
\label{a0b0}
&&
a_0 = b_0=- \frac{J_1(\ko a)}{H_1^{(1)}(\ko a)}\,,\\[5pt]
&&
\label{anbn}
a_n = b_n=- \frac{J_{\vert n\vert}(\ko a)}{H_{\vert n\vert}^{(1)}(\ko a)} \,,\quad n\ne 0\,.
\end{eqnarray}

\section{Discussion}
From \r{Qsi}, \r{Qei}, \r{Qsii}, \r{Qeii}, \r{a0b0}, and \r{anbn}, it follows that
\begin{equation}
Q_{sca}^{(i)}= Q_{sca}^{(ii)}=Q_{ext}^{(i)}= Q_{ext}^{(ii)}\,,
\label{qqq}
\end{equation}
because
\begin{equation}
\Big\vert \frac{J_n(\xi)}{H_n^{(1)}(\xi)}\Big\vert^2 = \Re\,\Big( \frac{J_n(\xi)}{H_n^{(1)}(\xi)}\Big)\,.
\end{equation}
The equality of extinction and  total scattering efficiencies for either case is an affirmation of the
nondissipative nature of nihility. The equality of efficiencies for both cases (i) and
(ii) emerges from the identity $a_n=b_n\,\forall n\in(-\infty,\infty)$ for nihility cylinders.

\begin{figure}[!ht]
\centering \psfull
\epsfig{file=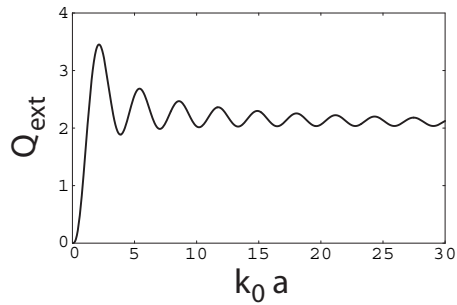,width=6cm }
\caption{Extinction efficiency of a nihility cylinder of cross--sectional radius
$a$ and infinite length when illuminated by a normally incident plane
wave of arbitrary polarization state.
}
\label{Fig1}
\end{figure}

An remarkable consequence of \r{qqq} is that the extinction and the total
scattering efficiencies of a nihility cylinder do not change with the polarization
state of the incident plane wave. In other words, if the incident plane wave
is arbitrarily polarized such that
\begin{equation}
\#E_{inc} = \left(A_z\hat{z}  +A_y\hat{y}\right)\, e^{-i\ko x}\,,
\end{equation}
the total scattering and the extinction efficiencies are independent of
the ratio $A_z/A_y$. The extinction efficiency is shown in Figure \ref{Fig1}
as a function of the normalized size parameter $\ko a$.

%%%%%%%%%%  Figure 1 begins %%%%%%%%%%%%

%%%%%%%%%%  Figure 1 ends  %%%%%%%%%%%%

Furthermore, on examining the scattering function
\begin{equation}
\#F_{sca}(\phi) = \lim_{\ko\rho\to\infty}\,(\ko\rho)^{1/2}\,\exp(-i\ko\rho)\,\#E_{sca}
(\rho,\phi)\,,
\end{equation}
it can be deduced  that the scattering
pattern $\vert \#F_{sca}(\phi)\vert$ is independent of the polarization state
of the incident plane wave. Again, this is because $a_n=b_n\,\forall n$.

The equality of scattering coefficients for
cases (i) and (ii) is a curious result,
at first glance. From \r{an} and \r{bn}, it can be shown that $a_n=b_n$ for all
$n$ if and only if $\eta_r^2=1$. On writing $\mur=\eta_r^2\epsr$, it becomes clear
that nihility is impedance--matched to free space (i.e., $\eta_r=1$). Indeed,
nihility is impedance--matched to any isotropic, homogeneous, dielectric--magnetic
medium (i.e, with both $\epsr\neq 1$ and $\mur\neq 1$), so that the results derived in Sec. \ref{spec} are very general. Parenthetically, we note that on repeating the 
exercise in Section 2 for obliquely incident plane waves led to expressions that could
not be unambiguously interpreted after the limiting procedure was implemented.

% ************* Closing part of the manuscript ****************

% ----- Acknowledgement:
% Activate if applicable
\Acknowledgement{Partial financial support from Center for the Integration of
Research, Teaching, and Learning funded by the US National
Science Foundation is acknowledged.}
%}

% ----- Appendix:
% Activate if applicable
% \begin{Appendix}      \end{Appendix}

% ----- References:
%\References{}{aeupap2e}{aeulit2e}
% in the case of explicit referencing (no crossreferencing) use instead:
% \References{*}{aeupap2e}{aeulit2e}

\end{document}